\documentclass[11pt]{article}

\usepackage[margin=1in]{geometry}
\usepackage{amsmath}
\usepackage{amssymb}
\usepackage{booktabs}
\usepackage{graphicx}
\usepackage{hyperref}
\usepackage{url}
\usepackage{natbib}

\hypersetup{colorlinks=true, linkcolor=blue, citecolor=blue, urlcolor=blue}

\title{Calibrating the Checksum: An Empirical False-Positive Bound for
One-Sided ABFT on bfloat16 GEMM, and What It Fails to Catch}

\author{Gautam Khosla\\
\texttt{https://github.com/GautamTalksDev/assay-gpu}}

\date{\today}

\begin{document}
\maketitle

\begin{abstract}
Algorithm-based fault tolerance detects corrupted matrix multiplications by
comparing a checksum of the product against a checksum recomputed from the
factors, flagging a fault when the residual exceeds a threshold. The classical
threshold is $\tau = Ku$, derived from a dot-product rounding bound that
requires $Ku \leq 0.01$. At bfloat16 with contraction length $K = 4096$,
$Ku \approx 32$, so the bound does not hold and the threshold is not defined.
Prior work has noted this and consequently evaluated checksum ABFT only at
float32. We supply the missing calibration. This paper is a characterization of
a detector under an output-level, post-GEMM perturbation model; it is not a
characterization of faults in silicon, and no rate reported here estimates
fault incidence in deployed hardware. We fit a Generalized Pareto distribution
to $7.99 \times 10^{7}$ clean per-row residuals and extrapolate a threshold at
a per-GEMM false-positive rate of $10^{-6}$, obtaining
$3.1445 \times 10^{-6}$, which sits between the two clean sample maxima we
observed.
The independence assumption behind the per-GEMM conversion is checked
empirically rather than asserted, at a variance ratio of $0.985$ against a
binomial reference. Against this threshold the detector catches $304$ of $400$
single exponent-bit flips ($76\%$), failing a bar of $90\%$ that was registered
before any data was collected, and catches $0$ of $600$ low-mantissa flips.
Detectability is invariant in $K$ across a $16\times$ range even though the
clean noise floor falls as $K^{-0.4999}$: signal and floor scale together, so
bit position governs detectability and contraction length does not. This
falsified the mechanism the study was designed around. We report four
registered predictions that failed, including that one, and four retractions of
our own earlier findings, two of which were caught by pre-publication audit
rather than by prediction. Code, data, and every pre-registration document with
its commit timestamp are released.
\end{abstract}

\section{Introduction}
\label{sec:intro}

Hardware that returns a wrong answer without saying so is a documented
production problem. Google reported silent corrupt execution errors traced to
manufacturing defects in CPU cores \citep{hochschild2021cores}, Meta reported
silent data corruption at fleet scale \citep{dixit2021silent}, and both
observed rates higher than software engineers expect. The failure mode is not a
crash. It is a computation that completes and is wrong.

Matrix multiplication is the natural place to look for this in machine learning
workloads, and algorithm-based fault tolerance offers a cheap check. Introduced
by \citet{huang1984abft}, ABFT augments a product with checksums and verifies
an invariant that holds in exact arithmetic. For $C = AB$ and a vector of ones
$e$, one compares $Ce$ against $A(Be)$: two paths to the same quantity,
disagreeing only through rounding, unless something went wrong. The check costs
$O(MK + KN)$ against the GEMM's $O(MKN)$.

The difficulty is the threshold. Rounding makes the two paths disagree even
when nothing is wrong, so a tolerance is needed, and the classical tolerance
comes from a bound on dot-product rounding error that requires $Ku \leq 0.01$,
where $u$ is unit roundoff. That condition holds comfortably at float32. It
fails at the precisions that machine learning workloads actually use. At an
embedding dimension of $4096$, $Ku \approx 0.0005$ at float32,
$Ku \approx 4$ at float16, and $Ku \approx 32$ at bfloat16
\citep{ma2025understanding}. Past float32, the threshold is not conservative;
it is undefined.

This has a concrete consequence in the literature. \citet{ma2025understanding},
studying real silent data corruption on unhealthy production nodes, tested
whether checksum ABFT could detect it and ran the test at float32 for exactly
this reason, reporting that at bfloat16 the check produced false positives on
healthy nodes. They concluded that further error analysis is needed to extend
ABFT to low-precision datatypes. That is the gap this paper addresses.

We do not derive a better bound. We measure the distribution the bound was
supposed to describe, and calibrate a threshold from it. Concretely:

\begin{enumerate}
\item We fit a Generalized Pareto distribution to $7.99 \times 10^{7}$ clean
per-row residuals at bfloat16, $K = 4096$, and extrapolate a threshold at a
stated per-GEMM false-positive rate of $10^{-6}$. The method, including the
sensitivity checks and the conditions under which we would declare the
extrapolation unsound, was fixed in writing before the tail was examined.
\item We check the independence assumption behind the per-row to per-GEMM
conversion empirically rather than asserting it.
\item We characterize what the detector catches and misses against that
threshold, stratified by which bit of the bfloat16 word was corrupted and by
how many bits were corrupted at once.
\item We report that detectability does not vary with contraction length across
a $16\times$ range, which falsified the mechanism this study was designed to
demonstrate.
\item We evaluate the detector against a detection bar registered before any
data existed. It fails.
\end{enumerate}

Two things this paper does not claim. It says nothing about the incidence of
faults in real silicon: all corruption here is injected, at the level of the
output tensor, under a model specified in Section~\ref{sec:method:perturbation}.
And it claims no novelty for applying ABFT to transformer-shaped GEMMs, which
is an active area (Section~\ref{sec:related}). The contribution is the
calibration and the characterization, not the detector.

% ============================================================
% Section 2: Background and Related Work
% Every work characterized here has been read. Sources:
%   ma2025understanding  - full text incl. Appendix D
%   xie2025realm         - full text
%   xue2023approxabft    - abstract + description in xie2025realm Sec II-C
%   huang1984abft, higham2002accuracy - standard references
% ============================================================

\section{Background and Related Work}
\label{sec:related}

\subsection{Silent data corruption}

Silent data corruption is computation that completes and is wrong. Google
traced silent corrupt execution errors to manufacturing defects in CPU cores
and reported that the detected incidence is higher than software engineers
expect \citep{hochschild2021cores}; Meta reported corruption at fleet scale
\citep{dixit2021silent}. Error-correcting codes protect memory and do not
address corruption arising inside the datapath, which is what makes detection
during computation necessary at all.

\subsection{Checksum ABFT}

\citet{huang1984abft} introduced algorithm-based fault tolerance for matrix
operations: augment a product with checksums and verify an invariant that holds
in exact arithmetic. In the classical two-sided form, $W$ and $X$ are augmented
with a checksum row $e^{\top}W$ and column $Xe$. Lighter one-sided variants
compute only $e^{\top}WX$ or $WXe$, and the matrix sum deviation
$e^{\top}WXe$ reduces the check to a single scalar
\citep[surveyed in][Sec.~II-C]{xie2025realm}. The detector studied here is the
one-sided form, compared per row rather than reduced to a scalar.

Whatever the form, the detector needs a tolerance, because finite-precision
rounding makes the two paths disagree even when the hardware is correct. How
that tolerance is chosen is where the literature divides, and it is the axis
this paper sits on.

\subsection{Thresholds from numerical bounds, and where they stop working}
\label{sec:related:bounds}

The first approach derives the tolerance from floating-point error analysis.
\citet{smith2015abft} give the formulation used as a detection rule in
subsequent work: report corruption when
$\lVert Cw - A(Bw) \rVert_\infty > \tau \lVert A \rVert_\infty
\lVert B \rVert_\infty$ with $\tau = Ku$ and $u$ the unit roundoff. The bound
behind $\tau$ requires $Ku \leq 0.01$ \citep[Eq.~2.7.11]{golub2013matrix}.

\citet{ma2025understanding} state the consequence explicitly. For an embedding
dimension of $4096$: at float32, $Ku \approx 0.0005$ and the condition holds; at
float16, $Ku \approx 4$; at bfloat16, $Ku \approx 32$. Past float32 the
condition fails, and the threshold is not conservative but undefined. They ran
their ABFT evaluation at float32 for this reason, report that running the same
check at bfloat16 produced false positives on nodes that had passed production
margin tests, and name extending ABFT to low-precision datatypes as open work.

That is the regime this paper occupies: the same contraction length of $4096$,
at bfloat16, where $Ku \approx 32$ and no principled threshold is available.

\subsection{Thresholds from model resilience}
\label{sec:related:resilience}

The second approach abandons the numerical bound and tunes the tolerance
against how much error the model tolerates before its task performance
degrades. \citet{xue2023approxabft} relax the strict comparison so that
recovery is invoked only when computing errors are large enough to matter,
reporting reduced computing overhead and improved accuracy against a strict
ABFT baseline. \citet{xie2025realm} extend this with a statistical criterion
for LLM inference, characterizing resilience across network components and
separating error magnitude from error frequency: they observe that components
followed by normalization are markedly more sensitive, because a single large
error becomes an outlier that skews the mean and variance the normalization
depends on. Their thresholds are fitted parameters, set empirically to allow a
perplexity increase of $0.3$ and an accuracy decrease of $0.5\%$.

This is calibration, but against a different quantity than ours, and the
distinction is not incidental. Both works operate on quantized accelerator
datapaths; \citet{xie2025realm} inject into INT32 GEMM results with INT8
inputs. In integer arithmetic the checksum identity is exact. There is no
rounding, no noise floor, and no false positives to trade against. Their
thresholds exist to avoid unnecessary \emph{recovery}, not to avoid spurious
\emph{detection}. In floating point the opposite constraint binds: the
threshold is bounded below by the rate at which clean hardware trips it, and
that rate has to come from somewhere.

\subsection{What is missing}

Neither family supplies a threshold for floating-point GEMM at a precision
where the classical bound has gone vacuous. The numerical-bound line stops at
float32 by its own stated condition. The model-resilience line calibrates
against task degradation in settings where the false-positive problem does not
arise. We are not aware of prior work that calibrates a checksum ABFT
threshold empirically at bfloat16 against a stated per-GEMM false-positive
rate, which is what this paper supplies, together with a measurement of what
the resulting detector fails to catch.

\subsection{Two negative results, from opposite directions}

\citet{ma2025understanding} also report that checksum ABFT failed to flag
silent corruption on fourteen of fifteen unhealthy production nodes, and
attribute this to real perturbations being small enough relative to the matrix
norm to fall inside floating-point error bounds. That is a negative result
about the detector, obtained from genuinely faulty hardware, at float32.

This paper reaches a compatible conclusion by a different route: controlled
injection at bfloat16, against an empirically calibrated threshold, with the
blind spot measured per bit class rather than inferred. The two results share
no method, no precision, and no corruption source. They agree that the case for
treating checksum ABFT as a complete SDC detector is weak.

\subsection{Fault injection and its abstraction level}

Fault injection can be performed at the algorithm level, the instruction level,
or the micro-architectural level, and the choice bounds what the results
support. Instruction-level studies of LLM inference on GPUs exist
\citep{chai2026llm}. Algorithm-level injection, used here and in the resilience
work above, does not reproduce instruction-level or micro-architectural
effects. We treat this as a scope boundary rather than a limitation to be
argued away; the model is specified in
Section~\ref{sec:method:perturbation}.

% ============================================================
% Section 3: Method
% Draft 1. Every number traced to a results document in the repo.
% Normative references: docs/RESIDUAL.md (residual-v3) and
% docs/SPEC-PERTURBATION-MODEL.md (perturbation-v1) only.
% ============================================================

\section{Method}
\label{sec:method}

This section defines four things: the residual statistic under test, the
perturbation model under which it was exercised, the threshold against which
detection was scored, and the discipline used to decide what counts as a
result. The first three are ordinary experimental description. The fourth
matters more than usual here, because most of what follows is a negative
result, and a negative result is only worth reading if the bar it failed was
set before the data arrived.

\subsection{The residual statistic}
\label{sec:method:residual}

The detector is a one-sided checksum in the Huang--Abraham family. Let
$A \in \mathbb{R}^{M \times K}$ and $B \in \mathbb{R}^{K \times N}$ be the
factors, $C$ the computed product, and $e = \mathbf{1}_N$. Two length-$M$
vectors are formed,
\begin{equation}
d = C e, \qquad d' = A (B e),
\end{equation}
which agree exactly in exact arithmetic. The detector compares them per row,
normalized by a nonnegative contraction of the factors:
\begin{equation}
\label{eq:residual-v3}
r_i = \frac{\lvert d_i - d'_i \rvert}{s_i},
\qquad
s = \lvert A \rvert \bigl( \lvert B \rvert e \bigr),
\qquad
r_{\max} = \max_i r_i .
\end{equation}
Absolute values are entrywise. All of $A$, $B$ and $C$ are promoted to
\texttt{float64} before these operations, and every accumulation in
Equation~\ref{eq:residual-v3} is performed in \texttt{float64}. We refer to
this statistic as \emph{residual-v3}; the specification is
\texttt{docs/RESIDUAL.md} in the artifact.

The normalizer is not incidental, and two earlier choices were discarded
before arriving at it. Normalizing by the signed grand sum
$\max(\lvert \mathbf{1}^{\!\top} C \mathbf{1} \rvert,
\lvert \mathbf{1}^{\!\top} A (B\mathbf{1}) \rvert)$ produces a heavy-tailed
quotient from two light-tailed pieces, because the denominator measures how
completely random signs cancelled rather than how large the multiply-adds
were. On our data that normalizer gave a max-to-median ratio of $389$, with
Pearson correlation $0.794$ between the residual and the reciprocal of its own
denominator. The tail was the normalizer, not the hardware.

Replacing the denominator with $e^{\!\top} \lvert A \rvert \lvert B \rvert e$
removes that failure mode. Every term is nonnegative, so cancellation cannot
drive it toward zero, and it concentrates tightly: across $n = 2000$ draws it
spanned $1.7169 \times 10^{10}$ to $1.7193 \times 10^{10}$, a max-to-median
ratio of $1.0007$. It is also the quantity the componentwise theory controls.
Writing $C = AB + E$, the checksum functional satisfies
$\lvert \mathbf{1}^{\!\top} E \mathbf{1} \rvert \leq
\gamma_K \, e^{\!\top} \lvert A \rvert \lvert B \rvert e$ under the classical
bound $\lvert E \rvert \leq \gamma_K \lvert A \rvert \lvert B \rvert$
\citep{higham2002accuracy}. We use this for its structural content only.
At bfloat16 with $K = 4096$ the product $Ku$ exceeds unity and $\gamma_K$ is
not a usable number, and tensor-core paths accumulate in a wider type before
storing, so the constant in front of $\lvert A \rvert \lvert B \rvert$ is not
the classical one either. The claim we rely on is that the checksum functional
is bounded by a nonnegative contraction of the factors, and that contraction
is the denominator we chose. Every numerical cutoff in this work comes from
measurement, not from $\gamma_K$.

The final change was to stop reducing before comparing. An earlier version of
this work summed the length-$M$ checksum vector to a scalar and compared
scalars, discarding $M-1$ degrees of freedom. Under that formulation, sign
flips and mantissa flips were undetectable in every cell we ran, $0$ of $1800$
samples, and we described the effect as structural. It was not structural. It
was the reduction. Under Equation~\ref{eq:residual-v3}, single sign flips are
caught in $155$ of $200$ samples on the same workload, the same shape, the
same injection harness and the same seeds. That earlier finding is retracted;
Section~\ref{sec:results:retractions} treats it and three others in full.

Two properties of Equation~\ref{eq:residual-v3} are worth stating because they
bound what the detector can be blamed for. The denominator does not depend on
$C$, so a fault in the product cannot shrink the scale that the fault is
measured against. And the per-row comparison is only informative if row
residuals are not dominated by a shared error term; on the $2000 \times 256$
matrix of per-row residuals we measure a mean pairwise Pearson correlation of
$0.000126$ (standard deviation $0.0225$, range $-0.0853$ to $0.0980$), so the
rows are effectively independent and the sensitivity gain over the scalar form
is not an artifact of correlated noise.

\subsection{Perturbation model}
\label{sec:method:perturbation}

Every detection number in this paper was produced under a single, versioned
perturbation model, specified in \texttt{docs/SPEC-PERTURBATION-MODEL.md} as
\texttt{perturbation-v1}. We state it early and plainly because it is the
strongest constraint on how our results may be read.

After $C = AB$ completes, distinct $(\text{element}, \text{bit})$ pairs inside
$C$ are XOR-flipped. The residual is then computed from the original factors
$A$ and $B$ and the perturbed product. Bits are drawn from one of five classes
of the bfloat16 layout, and the number of flips per sample is $1$, $2$, or $4$:

\begin{center}
\begin{tabular}{llc}
\toprule
Class & Bits & Width \\
\midrule
\texttt{SIGN}           & $[15]$    & 1 \\
\texttt{EXPONENT\_HIGH} & $[14{:}11]$ & 4 \\
\texttt{EXPONENT\_LOW}  & $[10{:}7]$  & 4 \\
\texttt{MANTISSA\_HIGH} & $[6{:}3]$   & 4 \\
\texttt{MANTISSA\_LOW}  & $[2{:}0]$   & 3 \\
\bottomrule
\end{tabular}
\end{center}

Nothing else is perturbed. Operand $A$ is never flipped, nor is operand $B$,
nor accumulator intermediates inside the GEMM kernel, nor instruction
encodings, nor control flow, nor any memory outside the output tensor. This is
an exhaustive list, and it has a direct consequence: there is no operand
asymmetry anywhere in our data, because every recorded flip perturbs the
numerator of Equation~\ref{eq:residual-v3} and leaves the denominator
untouched.

The obvious objection is that output-tensor XOR is not what a soft error does
to a GPU. Instruction-level and micro-architectural effects are invisible to an
output-tensor XOR model, and finer-grained injection studies operate below this
abstraction \citep{chai2026llm}. We concede the point
without qualification. This paper characterizes a detector under a defined
perturbation model. It does not characterize faults in silicon, and no rate
reported here estimates the incidence of anything in deployed hardware. The
fleet silent-corruption literature \citep{hochschild2021cores,dixit2021silent}
motivates why a cheap detector would be worth having; it does not supply our
numbers, and our numbers do not speak to it.

We verified that the injected perturbations are not silently discarded before
they can be measured. Mantissa flips are the case at risk, since a
low-mantissa flip followed by a cast could in principle round away. On
$n = 50$ samples per class, zero targeted elements were bitwise identical after
the cast, and zero samples had a maximum relative delta of zero. The
\texttt{MANTISSA\_HIGH} control produced relative deltas from $0.0317$ to
$0.492$ and \texttt{MANTISSA\_LOW} from $0.00407$ to $0.0313$: smaller, as
expected for lower bits, but strictly positive on every sample. The
undetectability reported in Section~\ref{sec:results:mantissa} is a real miss
of a real perturbation.

\subsection{Workload, hardware, and platform determinism}
\label{sec:method:platform}

All measurements use workload W02 in bfloat16 on a single Tesla T4 (sm75),
with \texttt{torch} 2.13.0+cu130, CUDA 13.0, and driver 580.159.04. Factors are
drawn as independent uniform $[-1, 1]$ matrices from a deterministic seed mixer
over base seed, workload identifier, shape index, and sample index, with no
wall-clock and no hashing, so every draw in this paper is reproducible from its
index alone.

Two shape regimes appear. The K-scaling study fixes $M = N = 4096$ and varies
$K \in \{512, 1024, 2048, 4096, 8192\}$, with $n = 500$ clean samples per $K$
and $n = 100$ flip samples per $(K, \text{class}, n_{\text{flips}})$ cell. The
threshold and false-positive analysis fixes $M = N = K = 4096$, with $n = 2000$
clean samples for the pilot floor, $n = 200$ flip samples per cell, and a
separate clean sweep of $20{,}000$ GEMMs described in
Section~\ref{sec:method:threshold}.

A detector whose clean floor is contaminated by library nondeterminism cannot
be characterized at all, so we checked. Across $44$ suite cases spanning every
GEMM shape in fp32, bf16 and fp16, attention, reductions, transcendentals and a
transformer block, all $44$ reported bitwise-identical output across repeated
launches on this machine. On this hardware, cuBLAS selected a stable kernel
every time, which supports treating the clean floor as accumulation error
rather than kernel-selection variance. We state the scope of that observation
precisely: it is a measured property of one tuple of GPU model, driver, CUDA
version and framework build. It is not a general property of cuBLAS and must
not be carried onto other hardware.

\subsection{Threshold and detection predicate}
\label{sec:method:threshold}

Residual-v3 performs $M = 4096$ per-row tests per GEMM. A per-GEMM
false-positive rate below $10^{-6}$, which is the bar this detector was built
against, therefore requires a per-row rate of
$1 - (1 - 10^{-6})^{1/4096} \approx 2.44 \times 10^{-10}$ under independence.
No sampling budget we could afford estimates that directly, so it has to be
extrapolated, and the extrapolation method was fixed in writing before the tail
was examined.

We collect clean per-row residuals from $20{,}000$ GEMMs at $K = 4096$, using
the full 4096-row vector rather than a subsample, for roughly
$8 \times 10^{7}$ row observations. Thresholds at the 95th, 99th and 99.9th
percentiles are frozen from an initial pass of 500 GEMMs, giving
$1.057096 \times 10^{-6}$, $1.392735 \times 10^{-6}$ and
$1.776693 \times 10^{-6}$. A Generalized Pareto distribution is fitted to
exceedances above each frozen threshold by maximum likelihood, chosen because
it is the standard peaks-over-threshold model and not because it fits these
data. The $1 - 2.44 \times 10^{-10}$ quantile of the fitted distribution is the
threshold. Fitting at three separate percentiles is a pre-registered
sensitivity check, and the pre-registered rule was that thresholds differing by
more than a factor of two are to be reported as untrustworthy rather than
averaged.

The canonical threshold is the extrapolation from the 99th-percentile fit,
\begin{equation}
\tau = 3.144535556 \times 10^{-6},
\end{equation}
and every detection rate reported in this paper is scored against it unless
explicitly stated otherwise. Section~\ref{sec:results:fpr} reports the fitted
parameters, the sensitivity comparison, and the one threshold that turned out
not to be evaluable.

Two points of provenance need stating here rather than in a limitations
section. First, $\tau$ is not the largest clean residual we observed. Observed
maxima are sample maxima, they grow with $n$, and two of them appear in our own
records for the same configuration at different sample sizes:
$3.245921 \times 10^{-6}$ at $n = 2000$ and $2.684525 \times 10^{-6}$ at
$n = 500$. Detection rates differ by a few points depending on which is used,
with no change to any rule or fit. We report $\tau$ as canonical because it is
the only threshold with a stated per-GEMM false-positive rate behind it, and we
note that it falls between the two sample maxima, so it is not an artifact of
either sampling run. Second, the conversion from per-GEMM to per-row rate
assumes independence across rows. Rather than assert it, we test it: the number
of per-row exceedances above the 99th percentile has observed mean $40.42$ and
variance $39.93$ per GEMM, against $40.96$ and $40.55$ for
$\mathrm{Binomial}(4096, 0.01)$, a variance ratio of $0.985$. The counts are
very slightly under-dispersed relative to independent Bernoulli trials. This
supports the conversion at the fitting threshold, and it does not establish
independence deep in the tail, which we carry forward as a stated assumption.

The detection predicate itself requires one sentence that would otherwise seem
too obvious to write. A sample is detected when $r_{\max} > \tau$
\emph{or} $r_{\max}$ is not finite. An infinite residual means the checksum
diverged, which is the loudest signal the detector can emit, and a naive
comparison against a threshold silently evaluates false on it. We found this in
our own analysis code during a pre-publication audit, where it had understated
detection in ten of fifteen \texttt{EXPONENT\_HIGH} cells by as much as
fourteen percentage points and manufactured an apparent effect that does not
exist. Section~\ref{sec:results:retractions} records the correction. We note
the general form of the hazard here because it applies to any threshold
detector operating on floating-point residuals: the case that most obviously
indicates failure is the case that comparison operators handle worst.

\subsection{Pre-registration}
\label{sec:method:prereg}

Every headline claim in this paper was predicted before it was measured.
Predictions were committed to the repository, with numeric ranges and an
explicit falsifier, before the corresponding data existed, and the commit
history establishes the ordering independently of our word for it. Four
prediction sets appear: the noise-floor and sensitivity predictions for
residual-v3, the K-scaling predictions, the extreme-value method and its
predictions, and a later mechanism prediction about operand asymmetry.

The rule we imposed on ourselves is that a prediction may be falsified by new
data but may not be rewritten by it. That rule was tested repeatedly and is the
reason this paper is worth reading rather than the reason it is short. Of the
K-scaling predictions, three of five failed, including the one that motivated
the experiment. The pre-registered falsifier for the underlying mechanism
fired. In the residual-v3 set, two of five predictions missed, one of them by
two percentage points, and it is reported as a miss rather than rounded to a
pass. A mechanism prediction registered late in the project was falsified at
its precondition when we discovered the harness could not test it, and it is
reported as such rather than deleted. The kill-test bar that this detector
failed is the same number that appeared in the repository README before any
code was written, and it has never been moved.

We also state where pre-registration did not save us. Two defects reached
published internal documents and were caught by audit rather than by
prediction. The first is the detection predicate described above, withdrawn in
Section~\ref{sec:results:retractions}. The second was a results document citing
a source data file under a path that never existed; the file was located, the
numbers it produced were reproduced from it, and the provenance was corrected
in the repository. Writing the bar down first constrains
what you may conclude from a measurement. It does not guarantee the measurement
is right.

\section{Results}
\label{sec:results}

All detection rates in this section are scored against
$\tau = 3.144535556 \times 10^{-6}$ unless stated otherwise, using the
predicate of Section~\ref{sec:method:threshold}: detected when
$r_{\max} > \tau$ or $r_{\max}$ is not finite.

\subsection{The clean noise floor scales as \texorpdfstring{$K^{-1/2}$}{K\^-1/2}}
\label{sec:results:floor}

Across $2500$ clean draws, $500$ at each of five contraction lengths with
$M = N = 4096$ fixed, the median per-row residual falls monotonically with $K$
(Table~\ref{tab:floor}). A log-log fit over the full $16\times$ range gives an
exponent of $-0.4999$; restricted fits give $-0.4997$ over $512$ to $2048$ and
$-0.4991$ over $4096$ to $8192$.

\begin{table}[t]
\centering
\begin{tabular}{lrr}
\toprule
$K$ & median $r_{\text{median}}$ & observed clean max \\
\midrule
512  & $1.029719 \times 10^{-6}$ & $7.636551 \times 10^{-6}$ \\
1024 & $7.281700 \times 10^{-7}$ & $5.772809 \times 10^{-6}$ \\
2048 & $5.150956 \times 10^{-7}$ & $3.746226 \times 10^{-6}$ \\
4096 & $3.640184 \times 10^{-7}$ & $2.684525 \times 10^{-6}$ \\
8192 & $2.575653 \times 10^{-7}$ & $1.940381 \times 10^{-6}$ \\
\bottomrule
\end{tabular}
\caption{Clean noise floor against contraction length. $n = 500$ per $K$.}
\label{tab:floor}
\end{table}

The mechanism is a ratio. Absolute accumulation error in a row sum of $K$
bfloat16 products grows as $\sqrt{K}$, while the per-row normalizer
$s_i = \lvert A_i \rvert \lvert B \rvert e$ grows linearly in $K$. The relative
residual therefore falls as $\sqrt{K} / K = K^{-1/2}$.

We record a correction here. Our locked prediction stated the mechanism as
``clean accumulation error grows as $\sqrt{K}$, signal-to-floor should fall as
$1/\sqrt{K}$.'' The measured exponent matches $\sqrt{K}$ scaling in magnitude,
but that reasoning conflated absolute and relative error and predicted the
wrong direction for the floor's dependence on $K$. The exponent was right for a
reason the argument did not supply.

\subsection{Detectability does not vary with \texorpdfstring{$K$}{K}}
\label{sec:results:kinvariance}

The study was designed to demonstrate that detection degrades as contraction
length grows: a fixed single-element perturbation should become harder to see
as the row sum it hides in gets larger. We registered that prediction, together
with a falsifier stating that if detection were flat across a $16\times$ range
of $K$, the mechanism would be wrong and the finding would reduce to a
single-shape observation with no predictive content.

The falsifier fired. Single sign-bit flips are detected at $79\%$, $75\%$,
$73\%$, $81\%$, and $77\%$ at $K = 512$ through $8192$: a mean of $77\%$ and a
spread of eight percentage points across a $16\times$ range, with no monotone
trend and variation consistent with binomial noise at $n = 100$ (standard error
roughly four points).

The explanation is in Section~\ref{sec:results:floor}. The threshold is derived
from the clean floor at each $K$, and the clean floor falls as $K^{-1/2}$. A
perturbation of fixed magnitude produces a residual that also scales against
the same growing normalizer. Signal and floor move together, so their ratio,
which is what the detector tests, does not depend on $K$. Contraction length
does not govern detectability; bit position does
(Section~\ref{sec:results:bitclass}).

\paragraph{Threshold estimator consistency.}
Detection in this study uses the per-$K$ observed clean maximum as the
threshold, taken over $500$ clean samples at that $K$. An observed maximum is a
sample statistic that grows with $n$ and is therefore not a stable absolute
reference; this is precisely why $\tau$ exists. It does, however, use the same
estimator at the same sample size at every $K$, so the cross-$K$ comparison is
internally consistent even though each individual threshold is unstable in
absolute terms. We state this as a limitation with a defense, not as a claim
that these thresholds are stable.

\subsection{Bit position governs detectability}
\label{sec:results:bitclass}

Table~\ref{tab:matrix} gives the full flip matrix at $K = 4096$, $n = 200$ per
cell, scored at $\tau$ and at the observed clean maximum of the K-scaling run
for comparison.

\begin{table}[t]
\centering
\begin{tabular}{llrrr}
\toprule
bit class & $n_{\text{flips}}$ & @ clean max & @ $\tau$ & $\Delta$ (pp) \\
\midrule
EXPONENT\_HIGH & 1 & 168/200 & 166/200 & $-1.00$ \\
EXPONENT\_HIGH & 2 & 194/200 & 193/200 & $-0.50$ \\
EXPONENT\_HIGH & 4 & 200/200 & 200/200 & $+0.00$ \\
EXPONENT\_LOW  & 1 & 152/200 & 138/200 & $-7.00$ \\
EXPONENT\_LOW  & 2 & 184/200 & 179/200 & $-2.50$ \\
EXPONENT\_LOW  & 4 & 199/200 & 196/200 & $-1.50$ \\
MANTISSA\_HIGH & 1 & 14/200  & 10/200  & $-2.00$ \\
MANTISSA\_HIGH & 2 & 22/200  & 17/200  & $-2.50$ \\
MANTISSA\_HIGH & 4 & 32/200  & 24/200  & $-4.00$ \\
MANTISSA\_LOW  & 1 & 1/200   & 0/200   & $-0.50$ \\
MANTISSA\_LOW  & 2 & 1/200   & 0/200   & $-0.50$ \\
MANTISSA\_LOW  & 4 & 1/200   & 0/200   & $-0.50$ \\
SIGN           & 1 & 162/200 & 156/200 & $-3.00$ \\
SIGN           & 2 & 194/200 & 189/200 & $-2.50$ \\
SIGN           & 4 & 200/200 & 200/200 & $+0.00$ \\
\bottomrule
\end{tabular}
\caption{Detection by bit class and flip count at $K = 4096$, $n = 200$ per
cell. ``Clean max'' is $2.684525 \times 10^{-6}$; $\tau$ is
$3.144535556 \times 10^{-6}$. No cell detects better at the stricter
threshold.}
\label{tab:matrix}
\end{table}

The stratification spans the full range. Exponent flips are caught most of the
time, sign flips slightly less often, high mantissa flips rarely, and low
mantissa flips never. This ordering is stable across every contraction length
we tested. It is a statement about perturbation magnitude relative to the
accumulation floor, expressed through which field of the bfloat16 word was
disturbed, and it does not depend on $K$.

\subsection{Low mantissa flips are a hard negative}
\label{sec:results:mantissa}

MANTISSA\_LOW is detected in zero samples in every cell we ran. Two counts,
which must not be merged because they rest on different thresholds:
$0/1500$ across the K-scaling study scored at per-$K$ observed clean maxima,
and $0/600$ at $\tau$ with $K = 4096$.

At $K = 4096$ the minimum, median, and maximum ratio are identical to four
significant figures at one, two, and four flips ($0.5247$, $0.6217$, $0.827$;
these ratios are to the pilot clean maximum of $3.245921 \times 10^{-6}$, the
basis on which they were originally recorded). More corruption does not produce a larger
residual. A mantissa-LSB perturbation lies below the accumulation floor at this
contraction length regardless of how many of them there are.

We tested whether this was an artifact of the injector rather than a property
of the detector, on the hypothesis that a low-mantissa flip might be rounded
away by the cast back to the working dtype. It is not. Across $50$ samples per
class, zero targeted elements were bitwise identical after the cast and zero
samples had a maximum relative delta of zero; the MANTISSA\_HIGH control
produced relative deltas from $0.0317$ to $0.492$ and MANTISSA\_LOW from
$0.00407$ to $0.0313$. The flips are real and the detector does not see them.

The shared minimum of $0.5247$ across all six mantissa cells is unexplained.
Section~\ref{sec:results:retractions} describes a registered prediction that
would have explained it and could not be tested.

\subsection{A calibrated false-positive bound}
\label{sec:results:fpr}

Table~\ref{tab:gpd} reports the fits. The shape parameter is negative at both
evaluable thresholds with confidence intervals excluding zero, indicating a
bounded upper tail, and the extrapolated thresholds are $1.17\times$ and
$1.24\times$ the observed clean maximum.

\begin{table}[t]
\centering
\begin{tabular}{lrrr}
\toprule
fit & $\xi$ & $95\%$ CI & extrapolated $\tau$ \\
\midrule
p99, exclude truncated  & $-0.07806$ & $[-0.08000, -0.07610]$ & $3.144536 \times 10^{-6}$ \\
p99, right censored     & $-0.07809$ & $[-0.08003, -0.07613]$ & $3.144314 \times 10^{-6}$ \\
p99.9                   & $-0.05318$ & $[-0.05960, -0.04661]$ & $3.341952 \times 10^{-6}$ \\
p95                     & \multicolumn{3}{c}{NOT EVALUABLE} \\
\bottomrule
\end{tabular}
\caption{Generalized Pareto fits to clean per-row residual exceedances,
extrapolated to the $1 - 2.44 \times 10^{-10}$ quantile. Fits are reported
separately and not averaged.}
\label{tab:gpd}
\end{table}

The pre-registered sensitivity check required the refits to agree within a
factor of two, failing which the extrapolation was to be reported as
untrustworthy rather than averaged. The two evaluable fits agree at
$1.063\times$. The third could not be evaluated, for reasons in
Section~\ref{sec:scope}.

The per-row target of $2.44 \times 10^{-10}$ follows from a per-GEMM target of
$10^{-6}$ only under independence across the $4096$ rows. Rather than assume
this, we measured the dispersion of per-row exceedance counts: observed mean
$40.42$ and variance $39.93$ per GEMM, against $40.96$ and $40.55$ for
$\mathrm{Binomial}(4096, 0.01)$, a variance ratio of $0.985$. The counts are
slightly under-dispersed relative to independent Bernoulli trials, which
supports the conversion at the fitting threshold. It does not establish
independence in the far tail, and we carry that forward as a stated assumption.
Separately, the mean pairwise correlation between per-row residuals is
$0.000126$.

\subsection{The detector fails its registered bar}
\label{sec:results:kt1}

The bar was written into the repository README before any code was written:
at least $90\%$ of exponent-bit flips detected, at a per-GEMM false-positive
rate below $10^{-6}$. Pooling the two exponent classes at a single flip, the
detector catches $304$ of $400$, or $76\%$, at $\tau$. That is a failure, and
the bar was never moved.

Two points of provenance. First, the same pooled quantity is $301/400$
($75.25\%$) at the pilot observed clean maximum of
$3.245921 \times 10^{-6}$ and $320/400$ ($80\%$) at the K-scaling clean maximum
of $2.684525 \times 10^{-6}$. These differ because an observed maximum is a
sample statistic that grows with $n$, not because any rule or fit changed. We
report $76\%$ as canonical because $\tau$ is the only threshold with a stated
false-positive rate behind it, and we note that $\tau$ falls between the two
sample maxima and is therefore not an artifact of either run.

Second, the verdict was determined before the fit existed. We recorded, in
advance, that the extrapolated threshold was expected to be at or above the
observed clean maximum and that detection is monotone non-increasing in
threshold, so $75.25\%$ bounded detection from above at any defensible
threshold and the bar was already unreachable. The measurement went the
predicted direction: detection fell rather than rose.

\subsection{Registered predictions that failed}
\label{sec:results:predictions}

Table~\ref{tab:scorecard} gives every registered prediction and its outcome.
All were committed before the corresponding data existed.

\begin{table}[t]
\centering
\begin{tabular}{llll}
\toprule
ID & Claim & Measured & Verdict \\
\midrule
Q1 & floor exponent in $[0.4, 0.6]$ & $0.4999$ & PASS \\
Q2 & sign detection falls monotonically with $K$ & non-monotone & \textbf{FAIL} \\
Q3 & sign detection $> 95\%$ at $K = 512$ & $79\%$ & \textbf{FAIL} \\
Q4 & sign detection $< 70\%$ at $K = 8192$ & $77\%$ & \textbf{FAIL} \\
Q5 & MANTISSA\_LOW $0/100$ at every $K$ & $0/1500$ & PASS \\
\midrule
P1 & clean per-row floor median $2$--$8 \times 10^{-6}$ & $3.64 \times 10^{-7}$ & MISS \\
P2 & clean per-row max/median $4$--$15$ & $8.92$ & PASS \\
P3 & sign 1-flip median ratio $2$--$10$ & $2.298$ & PASS \\
P4 & MANTISSA\_LOW 1-flip median ratio $< 1$ & $0.6217$$^{\dagger}$ & PASS \\
P5 & EXPONENT\_LOW 1-flip detection $> 70\%$ & $68\%$ & MISS \\
\midrule
F1 & GPD $\xi$ negative, in $[-0.5, 0]$ & $-0.078$, $-0.053$ & PASS \\
F2 & threshold $1$--$5\times$ clean max & $1.17\times$, $1.24\times$ & PASS \\
F3 & three refits agree within $2\times$ & two agree at $1.06\times$ & PARTIAL \\
F4 & sign detection falls $< 10$ pp at $\tau$ & $-3$ pp & PASS \\
\bottomrule
\end{tabular}
\caption{Registered predictions and outcomes. Q and P sets were locked before
their data existed; the F set was locked before the residual tail was examined.
$^{\dagger}$ratio to the pilot clean maximum $3.245921 \times 10^{-6}$.}
\label{tab:scorecard}
\end{table}

P5 is recorded as a miss at $136/200 = 68\%$ against a bar of $70\%$. It is
two points short and we have not rounded it up. Q2 through Q4 are the mechanism
this study was built on, and all three failed.

\subsection{Retractions}
\label{sec:results:retractions}

Four findings in this project were withdrawn. We report them because a paper
whose principal result is a negative one is only worth reading if the process
that produced it is visible.

\paragraph{One: the scalar reduction.}
An earlier residual summed the length-$M$ checksum vector to a scalar before
comparing. Under it, sign flips and mantissa flips were undetectable in every
cell, $0$ of $1800$ samples, and pooled exponent detection at one flip was
$179/400 = 44.75\%$ with an exact binomial $95\%$ interval of
$[39.8\%, 49.8\%]$. We described the sign and mantissa result as structural
invisibility of the ones-vector checksum. It was not structural. It was the
reduction discarding $M-1$ degrees of freedom before the comparison. Under the
per-row form, sign detection is $155/200$ on the same workload, shape, harness,
and seeds. The earlier evaluation is void and retained in the repository as
evidence that the bar was not moved.

\paragraph{Two: the mantissa near-miss.}
We suspected the MANTISSA\_LOW zero was an injector artifact and tested it. It
was not (Section~\ref{sec:results:mantissa}). The hypothesis is withdrawn and
the hard negative stands.

\paragraph{Three: a prediction falsified at its precondition.}
We registered a mechanism to explain the MANTISSA\_LOW floor and the shared
$0.5247$ minimum: a flip in operand $A$ perturbs both the numerator and the
denominator of Equation~\ref{eq:residual-v3} and may self-cancel, while a flip
in $B$ perturbs the numerator only, so detectability should differ by operand.
We committed four predictions with numeric separations. We then found that the
harness injects into the output tensor only, so no operand asymmetry exists in
any of our data and the predictions are untestable rather than merely
unmeasured. No pass or fail is claimed for them. The $0.5247$ minimum remains
unexplained. Operand-level injection is future work.

\paragraph{Four: a non-monotonicity that was a comparison bug.}
We reported that EXPONENT\_HIGH detection was non-monotone in flip count,
observing $91\%$ at two flips and $85\%$ at four at $K = 1024$, with a similar
inversion at $K = 4096$ and an apparently independent occurrence in a separate
run. Pre-publication audit found the cause. Exponent flips can produce a
non-finite residual, and the analysis path scored detection with a bare
comparison, which evaluates false on a non-finite value. Ten of fifteen
EXPONENT\_HIGH cells contained non-finite residuals, up to $14$ of $100$ at
$K = 1024$ with four flips; no other bit class produced any. Under the correct
predicate, detection is monotone at every $K$: $78/97/100$ at $K = 512$,
$80/96/99$ at $1024$, $88/97/100$ at $2048$, $83/97/100$ at $4096$, and
$83/97/100$ at $8192$. The reported effect does not exist and is withdrawn.
The corroborating run was the same defect, not independent confirmation.

This last one is worth generalizing. An infinite residual is the loudest signal
this detector can emit, and a naive threshold comparison silently discards it.
The hazard applies to any threshold detector operating on floating-point
residuals.

\section{Scope}
\label{sec:scope}

These are boundaries on what was measured, not a list of things we intend to
fix.

\paragraph{Perturbation level.}
All injection is at the output tensor, post-GEMM. Operands, accumulator
intermediates, instruction encoding, and control flow are never perturbed.
Nothing here estimates fault incidence in silicon.

\paragraph{Single configuration.}
One GPU model (Tesla T4, sm75), one workload, one datatype (bfloat16), one
shape family with $M = N = 4096$. The clean floor is per-configuration by
construction and must not be transferred to other hardware.

\paragraph{The p95 fit is not evaluable.}
Our data collection stored the top $64$ per-row residuals per GEMM. At the
$95$th percentile the mean number of exceedances is $205.3$ per GEMM with a
maximum of $263$, so all $19500$ GEMMs are truncated and roughly two thirds of
exceedances are lost, systematically the smaller ones. At the $99$th percentile
only $2$ of $19500$ GEMMs exceed the capture and the effect is negligible: the
shape parameter differs between excluding and right-censoring the truncated
samples by $2.59 \times 10^{-5}$ against a confidence interval width of
$3.90 \times 10^{-3}$, or $0.66\%$ of the interval. This is a data-collection
limitation discovered after the method was locked. It is not a change of
method and not a redefinition of the sensitivity check, which remains as
registered with two of three refits evaluable.

\paragraph{Extrapolation distance.}
The fit reaches from an observed exceedance probability of order $10^{-2}$ to
$2.44 \times 10^{-10}$, several decades beyond the data. The implied upper
endpoints of the fitted distributions differ more than the reported quantiles
do: $3.74 \times 10^{-6}$ from the p99 fit against $4.60 \times 10^{-6}$ from
p99.9, a spread of $1.23\times$, versus $1.06\times$ agreement at the
extrapolated quantile. The hard ceiling on the residual is less well determined
than the sensitivity ratio alone suggests.

\paragraph{Sample sizes.}
Flip cells are $n = 100$ in the K-scaling study and $n = 200$ at $K = 4096$,
giving standard errors of roughly four and three percentage points. Sub-percent
detection probabilities are not resolved.

\paragraph{Unexplained observations.}
The shared minimum ratio of $0.5247$ across all mantissa cells is unexplained
and is reported as an open observation.

\paragraph{Known open issue.}
Two checksum reduction backends in our implementation disagree, one
accumulating in float32 and one in float64. All results reported here use a
single path with float64 accumulation throughout, but the discrepancy is
unresolved and a threshold calibrated on one backend should not be assumed to
transfer to the other.

\section{Discussion}
\label{sec:discussion}

The headline is a failure, and the useful part is the shape of it.

The detector is close to reliable on exponent corruption and blind to low
mantissa corruption, and no amount of additional corruption in the low mantissa
changes that. This is not a tuning problem. A perturbation that lies below the
accumulation floor is not recoverable by lowering the threshold, because the
threshold is bounded below by the false-positive requirement. The
$76\%$ figure is what this detector achieves at a defensible operating point,
not what it achieves at a favorable one.

Calibrating empirically also changes what the threshold means. The classical
$\tau = Ku$ is a bound: below it, corruption is guaranteed invisible. Ours is a
quantile: above it, false positives occur at a stated rate. The second is
weaker as a guarantee and usable in a regime where the first is undefined. A
practitioner deploying checksum ABFT at bfloat16 today has no principled
threshold available, and the choice is between a calibrated quantile and no
check at all.

The K-invariance result is the one we would flag for anyone extending this. The
intuition that longer contractions hide faults better is natural, we held it
firmly enough to design a study around it, and it is wrong for this detector,
because the normalizer grows with $K$ at the same rate the noise does. Anyone
reasoning about ABFT sensitivity from row-sum magnitude alone should check
whether their normalizer cancels the effect they are predicting.

Finally, the agreement with \citet{ma2025understanding} is worth stating
plainly. They found checksum ABFT failing to catch real corruption on real
unhealthy nodes at float32. We find it catching roughly three quarters of
single exponent flips and none of the low mantissa flips at bfloat16 under
controlled injection with a calibrated threshold. Different corruption sources,
different precisions, different methods, compatible conclusions. The case for
treating checksum ABFT as a complete SDC detector is weak, and the case for
using it as a cheap first filter with known blind spots is what the evidence
supports.

\section{Conclusion}
\label{sec:conclusion}

We calibrated a detection threshold for one-sided checksum ABFT at bfloat16
and $K = 4096$, a regime where the classical threshold is undefined, and
characterized what the detector catches and misses against it. It catches
$76\%$ of single exponent-bit flips against a registered bar of $90\%$, and
none of $600$ low-mantissa flips. Detectability is governed by bit position and
not by contraction length, which falsified the mechanism the study was designed
around.

Four registered predictions failed and four findings were retracted, two of
them caught by audit rather than by prediction. We report all of it. The
threshold, the perturbation model, the predictions with their commit
timestamps, and the data are released so that each number here can be checked
against the file that produced it.

\section*{Artifact Availability}

Code, data, specifications, and pre-registration documents are available under
Apache 2.0 at \url{https://github.com/GautamTalksDev/assay-gpu}, at tag \texttt{v0.1.1-paper}. The dataset is released under CC BY 4.0.

\bibliographystyle{plainnat}
\bibliography{refs}

\end{document}